\documentclass{osa-article}

\journal{osac}

\articletype{Research Article}
\usepackage{siunitx}
\usepackage{textgreek}
\begin{document}

\title{Atomic line versus lens cavity filters: A comparison of their merits}

\author{Clare R. Higgins,\authormark{1,*} Danielle Pizzey,\authormark{1}, Renju S. Mathew,\authormark{1} and Ifan G. Hughes\authormark{1}}

\address{\authormark{1}Joint Quantum Centre (JQC) Durham-Newcastle, Department of Physics, Durham University, South Road, Durham, DH1 3LE, UK}

\email{\authormark{*}clare.r.higgins@durham.ac.uk} 



\begin{abstract}
We present a comparison between lens cavity filters and atomic line filters, discussing their relative merits for applications in quantum optics. We describe the design, characterization and stabilization procedure of a lens cavity filter, which consists of a high-reflection coated commercially available plano-convex lens, and compare it to an ultra-narrow atomic band-pass filter utilizing the D$_{2}$ absorption line in atomic rubidium vapor. We find that the cavity filter peak transmission frequency and bandwidth can be chosen arbitrarily but the transmission frequency is subject to thermal drift and the cavity needs stabilization to better than a few mK, while the atomic filter is intrinsically stable and tied to an atomic resonance frequency such that it can be used in a non-laboratory environment.
\end{abstract}

\section{Introduction}
Optical filters are used in a variety of applications for isolating a signal frequency from unwanted background noise. The best commercially available thin-film interference band-pass filters typically have transmission bandwidths of a few nanometres, where the transmission bandwidth is defined as the full-width-at-half-maximum (FWHM). In some research fields, in particular quantum optics where it is necessary to distinguish single photons from high background counts at similar wavelengths \cite{Lvovsky:12,Sinclair:14,Zielinska:14, Tan_2014}, narrow-band optical filters are required: these filters can have \mbox{sub-nm} transmission bandwidths, while still retaining high on-peak transmission and off-peak extinction. Uses include the demonstration of quantum teleportation \cite{Bouwmeester:97}, quantum memory \cite{Hosseini:11, England:15} and quantum information processing \cite{Yukawa:13}, as well as in other fields of atomic physics such as atom trapping \cite{Schlosser:02}. Examples of narrow-band filters include atomic line filters and cavity filters; these shall be the focus of this discussion.

Atomic line filters are often used in atmospheric LIDAR \cite{Fricke-Begemann:02, Popescu2004, Gong2015}, optical communications \cite{Bloom:93}, and laser frequency stabilization \cite{Keaveney:16,Miao:11}. These filters consist of an atomic vapor cell placed between two crossed polarizers and subject to a magnetic field which causes the polarization of light to be rotated as it traverses the cell \cite{Faraday:1846}, leading to transmission through the second polarizer.  For an atomic medium, polarization rotation only occurs near atomic resonances (which are intrinsically narrow), producing a narrow filter \cite{Keaveney:18,Abel:09}. Filters have been demonstrated in different atomic species, including Cs \cite{Menders:91, Rotondaro:15,FADOF1GHz},  Rb \cite{Dick:91} and Na \cite{Chen:93, Kiefer2014}. The most commonly used atomic filter geometries have been the Faraday geometry, where the magnetic field, $\vec{B}$, and light propagation direction, $\vec{k}$, are parallel, and Voigt geometry where $\vec{B}$ and $\vec{k}$ are perpendicular. However it is also possible to construct a filter with an arbitrary angle between $\vec{B}$ and $\vec{k}$, which is more computationally complex. The transmission spectrum behaves non-trivially as a function of temperature, magnetic field, and polarizer angle for a given cell length. Therefore it is advantageous to use an accurate model of the filter spectrum to find optimum operating parameters; we used a computational model, ElecSus \cite{Zentile2015,Keaveney2017}.

Cavity filters consist of two high-reflectivity (HR) dielectric coated surfaces, which are separated by a predetermined length \cite{Hadley:47}. Any light entering the cavity through the first surface will only exit at the second surface when it is resonant with the cavity and the standing wave condition is met \cite{f2f}, resulting in a periodic set of transmission peaks and high extinction elsewhere.  

 We have built and investigated two narrow-band band-pass filters: a lens cavity filter \cite{lvovsky} and an atomic line filter in Rb vapor \cite{Keaveney2017}. Each has advantages and disadvantages---in this article we present a study comparing the two.

\section{Lens Cavity filter}
\label{sec:lens}

We implement a monolithic cavity filter, as proposed in \cite{lvovsky}, where a spherical high-reflection ($R$\,$\sim$\,99\,\%) coated plano-convex lens is used as the Fabry-P\'erot cavity. This setup produces transmission peaks with widths of the order 70~MHz, where the central frequency is tunable with temperature. We chose to investigate this cavity design, over the well-known design that consists of two HR coated mirrors attached to a spacer of a given length, because it is intrinsically stable and requires no locking of mirror positions. The plano-convex geometry also provides spatial mode filtering, and allows a higher cavity finesse than a planar etalon cavity \cite{lvovsky}. The cavity quality is governed by finesse, and ideal finesse of the cavity is given by $\mathcal{F} = \left({\pi \sqrt{R}}\right) / \left({1-R}\right)$.
The achievable finesse is limited by surface defects and the mismatch between the wavefront and the surface. Due to diffraction, it is not possible to mode match wavefronts to two separated planar cavity mirrors. Choosing a larger beam diameter to compensate increases the effect of surface defects, limiting the finesse of a flat etalon to~100. The spherical mirrors of a Fabry-P\'erot are much more forgiving, allowing a finesse up to $\num{1e5}$. \cite{Maunz2004}

The required temperature stability of the cavity is determined by the desired frequency stability and the shift of resonant frequency, $v$, with temperature, $T$, given by
\begin{equation}
\frac{\text{d}v}{\text{d}T} \approx -\left(\alpha + \frac{1}{n}\frac{\delta n}{\delta T}\right) v,
\label{eqn:tempstab}
\end{equation}
where $\alpha = \SI{7.1e-6}{\per\kelvin}$ and $n = 1.51$ are the thermal expansion coefficient and refractive index of Schott N-BK7 Glass and $\delta n/\delta T$ is calculated from the Sellmeier function \cite{schott}. The change in refractive index due to frequency is negligible in comparison to the other terms. The filter design parameters are: reflectivity, $R$; the radius of curvature of the convex face, $r$; and the thickness (or length), $L$. 
The transmission of a Fabry-P\'erot cavity, $\mathcal{T}$, is given by \cite{f2f}
\begin{equation}
\mathcal{T} = \left(\frac{T}{1-R}\right)^2 \bigg(1+\frac{4R}{({1-R})^2} \sin^2{\frac{\delta}{2}}\bigg)^{-1},
\label{eqn:FP}
\end{equation}
where $\delta$ is the frequency dependent per round trip phase shift, and $T$ is the transmission at the mirror, which may not equal $1 - R$ due to round trip losses. When $R$ is close to 1, the extinction ratio is 
\begin{equation}
\mathcal{T}_\text{max}/\mathcal{T}_\text{min} \approx \frac{4}{({1-R})^2} \approx \bigg(\frac{2\mathcal{F}}{\pi}\bigg)^2,
\end{equation}
so a required extinction ratio sets the reflectivity. The bandwidth of the transmission peaks (\textDelta$v$) is given by \textDelta$v = \text{FSR}/\mathcal{F}$, so a required bandwidth sets the free spectral range (FSR) of the cavity. The length is set via $\text{FSR} = c/{2nL}$, where $c$ is the speed of light. The spatial filtering requirements determine $r$; in the case that $L \ll r$, adjacent transverse modes are separated by 
\begin{equation}
\Delta v_\perp = \frac{\text{FSR}}{\pi} \sqrt{L/r}, 
\label{eqn:modes}
\end{equation}
 and $r$ can be chosen to ensure that no significant subsidiary modes transmit at a frequency of interest.

\begin{figure}[htbp]
\centering
\includegraphics[width=0.85\linewidth]{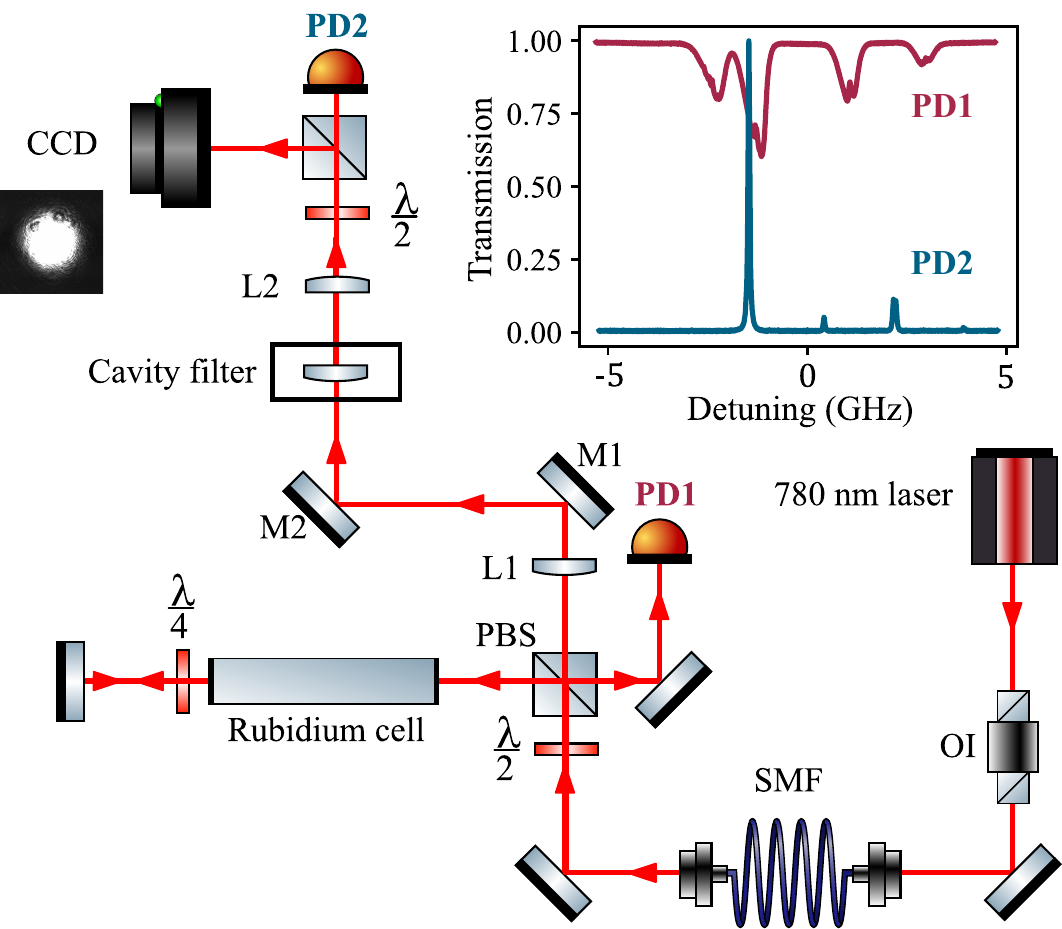}
\caption{Experimental setup for cavity filter characterization. 780~nm laser light is passed through an optical isolator (OI) and a single mode fiber (SMF), producing a beam with waist \SI{0.5}{mm} and then split on a polarizing beam splitter (PBS) cube. One arm double passes through a natural abundance rubidium vapor cell to provide an atomic frequency reference (red trace) on a photodiode (PD1). The other beam continues through a mode-matching lens (L1) and is steered by mirrors (M1 and M2) into the lens cavity filter. The beam is recollimated by lens (L2), and split with another PBS, allowing the transmission (blue trace) to be monitored on a photodiode (PD2) and the output mode to be imaged on a CCD camera.}
\label{fig:etalonsetup}
\end{figure}

The off-the-shelf lenses were purchased from, then coated by Lambda Research Optics Inc, with $R$\,=\,(99.0\,$\pm$\,0.5)\,\% for wavelengths in the range 740--860~nm. We chose five different cavity lengths, in the range 2.4--8.0~mm, resulting in cavity bandwidths ranging from 40--128~MHz. 
All have radius of curvature $r$\,=\,\SI{40.0}{mm}.

Using Eq.\,(\ref{eqn:tempstab}), with $\alpha = \SI{7.1e-6}{\per\kelvin}$, $n = 1.51$, $\delta n/\delta T = \SI{2.54e-6}{\per\kelvin}$ and $v = \text{c}/\SI{780}{nm}$ gives a frequency shift with temperature of $\text{d}v/\text{d}T = \SI{-3.4}{\giga\hertz\per\kelvin}$, meaning a temperature stability of \SI{3}{\milli\kelvin} is required for frequency stability of 10~MHz. This number is independent of the length of the cavity. We chose BK7 glass because a range of lenses matching our specification were commercially available, and the temperature stabilization required is achievable. A glass with a lower thermal expansion coefficient or thermal refractive index change could be chosen if a higher frequency stability is required. To achieve this temperature stability, we mount the lens in a lens tube, which screws into a stainless steel block. This is thermally contacted to a peltier and thermistor, and is further encased in a teflon cover. The cool side of the peltier is contacted to a large aluminium block mounted on the optical bench, providing a large heat sink. The temperature is controlled with a Koheron TEC100L temperature controller.

The experimental setup used to characterize the performance and stability of the cavity filter is shown in Fig. \ref{fig:etalonsetup}. The cavity must be aligned to couple the correct mode (TEM$_{00}$). It is necessary to mode match into the fundamental mode of the cavity using a lens (L1), which is selected to match the curvature of the wavefronts with the spherical (front) surface of the cavity, focussing the beam at the planar surface. The beam waist of the fundamental cavity mode at wavelength $\lambda$ is  \cite{mccraethesis}
\begin{equation}
w_0 = \bigg(\frac{\lambda n L}{\pi}\bigg(\frac{r}{nL}-1\bigg)^{1/2}\bigg)^{1/2}.    
\end{equation}
The focal length of the mode-matching lens required is given by $f = \frac{w_1 w_0 \pi}{\lambda}$ \cite{f2f}, where $w_1$ is the initial beam waist. Two steering mirrors (M1 and M2) after the mode-matching lens are used to optimize alignment into the cavity, giving full control of the x-y position and angle of the beam when it enters the cavity. The alignment is monitored mainly on the transmission spectrum, with the mode image used to identify which is the TEM$_{00}$ mode to maximize.

\begin{figure}[htbp]
\centering
\includegraphics[width=\linewidth]{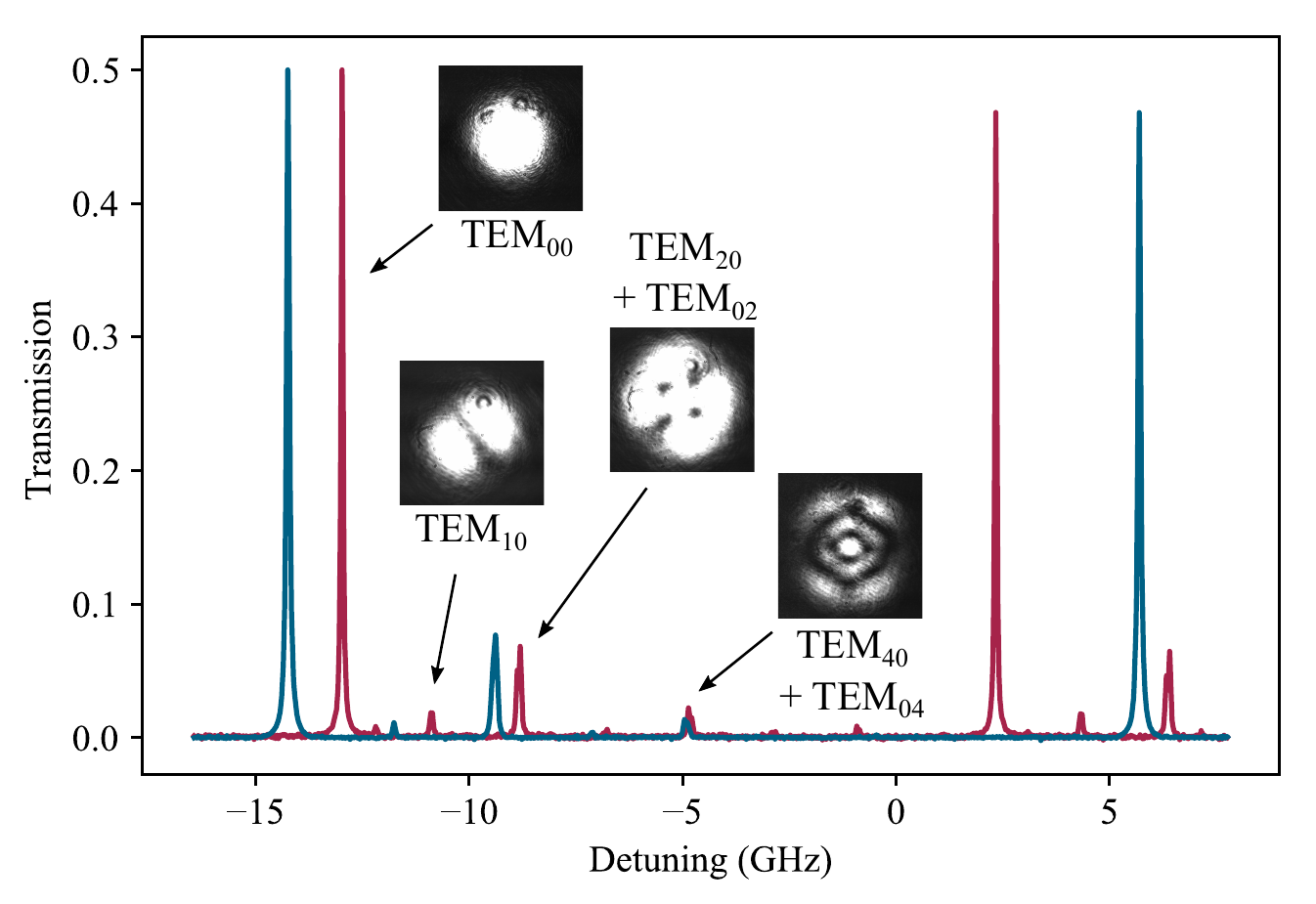}
\caption{Normalized transmission spectra of cavities of length 5.0~mm (blue) and 6.5~mm (red), with CCD images of the transmitted modes. The TEM$_{00}$ peaks are separated by (19.9~$\pm$~0.1)~GHz (5.0~mm) and (15.3~$\pm$~0.1)~GHz (6.5~mm). Higher order modes appear in between, with TEM$_{10}$, TEM$_{20}$ + TEM$_{02}$ and TEM$_{40}$ + TEM$_{04}$ (pictured) clearly visible. Spacing between the modes is 2.2~GHz (5.0~mm) and 1.9~GHz (6.5~mm) as expected.}
\label{fig:fsr-modes}
\end{figure}

\begin{figure}[htbp]
\centering
\includegraphics[width=\linewidth]{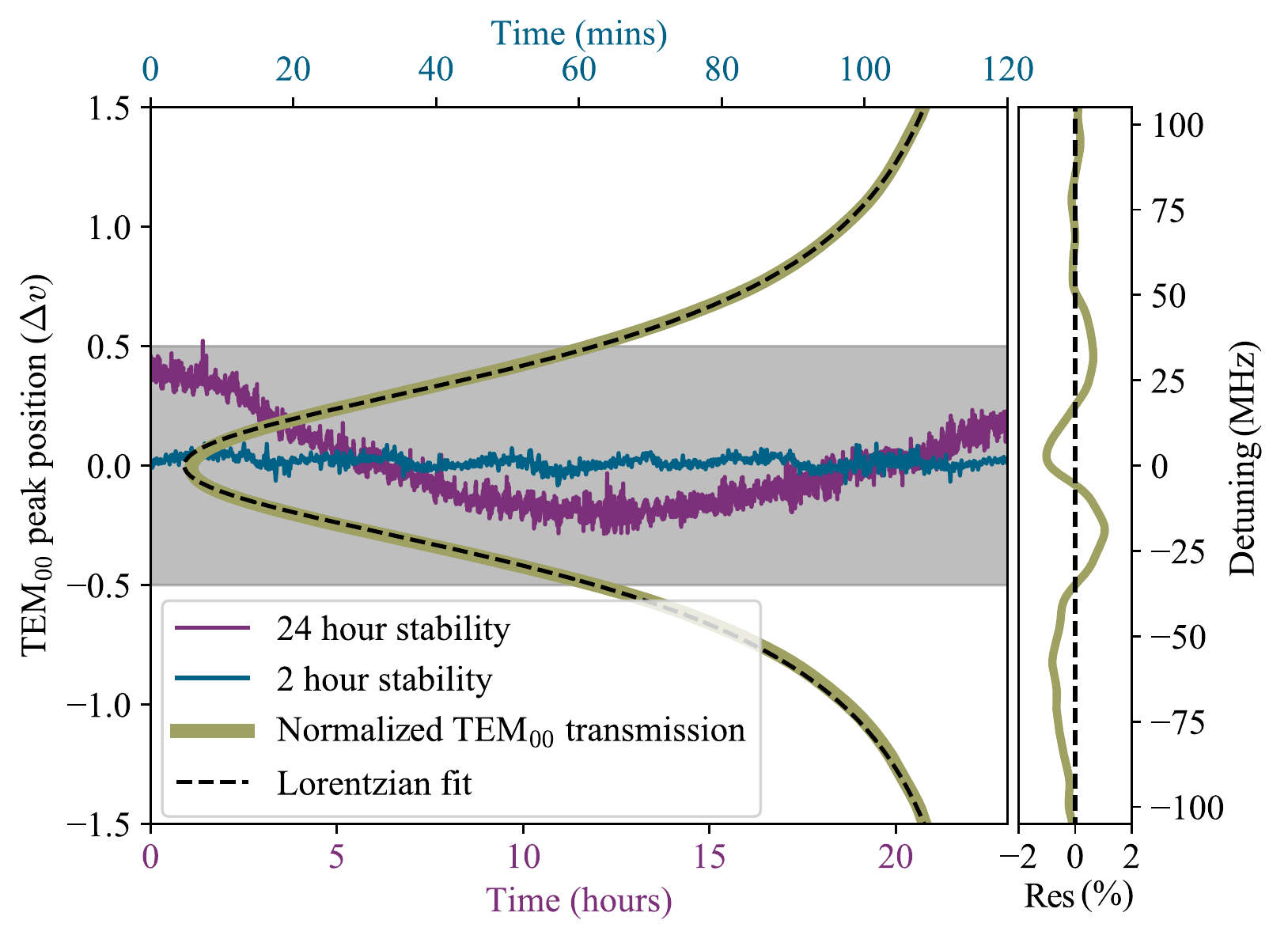}
\caption{The stability of the 5.0~mm cavity over a 2 hour (blue) and 24 hour (purple) period. The TEM$_{00}$ mode of the cavity is overlaid (gold), with a Lorentzian fit (black), and residuals. The grey shaded area shows one FWHM bandwidth (\textDelta$v$). Over 2(24) hours the cavity is stable to 0.1(0.7) \textDelta$v$.}
\label{fig:stability}
\end{figure}

Fig. \ref{fig:fsr-modes} shows a transmission spectrum across one FSR of the 5.0~mm  and 6.5~mm lens cavities. The maximum transmission of both cavities was measured to be 50\,\%, with extinction of 20 dB over all frequencies away from the TEM$_{00}$ mode. Maximum transmission is not 100\,\% largely due to imperfect mode-matching into the cavity. The spatial filtering properties of the filter are visible with subsidiary modes transmitting at different wavelengths, determined by Eq.\,(\ref{eqn:modes}).  We fit the TEM$_{00}$ mode of the 5.0\,mm cavity to a Lorenztian (gold solid and black dashed lines in Fig. \ref{fig:stability}), finding excellent agreement with residuals less than 1\%. This is expected as in the high finesse limit the transmission peaks given by Eq.\,(\ref{eqn:FP}) become Lorenztian \cite{f2f}. From this a width of (70\,$\pm$\,1)\,MHz is extracted. This is close to the expected value of 64\,MHz, and gives the actual reflectivity of the coating as (98.9\,$\pm$\,0.1)\,\%. A figure of merit often used to characterize the performance of optical filters is equivalent noise bandwidth, defined as $\text{ENBW} = \int T(\omega) \text{d}\omega / T (\omega_s)$, where $T$ is transmission, $\omega$ is angular optical frequency and $\omega_s$ is the frequency of maximum transmission. However, for applications that require a high peak transmission and narrow bandwidth, a better figure of merit is $\text{FOM} = T (\omega_s)/\text{ENBW}$ \cite{Keaveney:18}. FOM rewards filters with high maximum transmission and low overall transmission. As cavity peaks repeat every FSR the FOM for the lens cavity filter, if evaluated over all frequencies, is zero. However if we limit the calculation to one FSR, we obtain values of (6.5\,$\pm$\,0.1)\,$\text{GHz}^{-1}$ and (7.3\,$\pm$\,0.1)\,$\text{GHz}^{-1}$ for the 5.0~mm and 6.5~mm cavities respectively.

We determine the filter temperature stability by tracking the frequency of the TEM$_{00}$ peak relative to a sub-Doppler rubidium spectral line. This allows us to passively monitor the cavity over many hours or days, sampling every second, and accounts for any laser frequency drift that may occur. Fig. \ref{fig:stability} shows the stability of the 5.0~mm cavity over a period of 24 hours, where there is a long term drift of 0.7\,\textDelta$v$ which we attribute to fluctuations in laboratory temperature. Also shown is the drift in a 2 hour window, during which the cavity peak is stable to 0.1\,\textDelta$v$.

We experimentally determine the resonant frequency change with temperature, $\text{d}v/\text{d}T$ (Eq.\,(\ref{eqn:tempstab})), to be (-3.36\,$\pm$\,0.06)\,\SI{}{\giga\hertz\per\kelvin} in agreement with the expected value of \SI{-3.4}{\giga\hertz\per\kelvin}. As the transmission spectrum repeats every FSR, the maximum temperature change required is that to shift the modes by half the FSR: 9.9\,GHz for the 5.0~mm cavity. We characterize the response time of the 5.0~mm cavity to a change in temperature set point, and find that for smaller temperature changes (up to 2\,K, 6.6\,GHz) peak movement is well fitted to an exponential, with a 1/e time constant of 1.5~mins. This is expected, and is a signature of the thermal capacity of the lens. When the temperature change is larger than this the TEC board reaches the current output limit so peak movement is initially slower than exponential. For a 3.3\,GHz shift (1.0\,K), the time taken for the cavity to become stable to within \textDelta$v$ is 10~mins; for an 11.1\,GHz shift (3.4\,K) this rises to 20~mins. 

Any birefringence in the glass of the cavity, caused by stress in the material, will induce a shift in the transmission frequency  with changing polarization, as noted in \cite{lvovsky}. Our mounting method minimizes stress across the lens and we measure a negligible frequency shift with rotation of incoming linear polarization. Testing two lenses, we measure maximum shifts of 5\,MHz and 10\,MHz respectively, which is on the order of the fluctuations due to temperature instability, as seen in Fig. \ref{fig:stability}. There is also no significant change for circularly polarized input light.
\section{Atomic line filter}

\begin{figure}[htbp]
\centering
\includegraphics[width=0.8\linewidth]{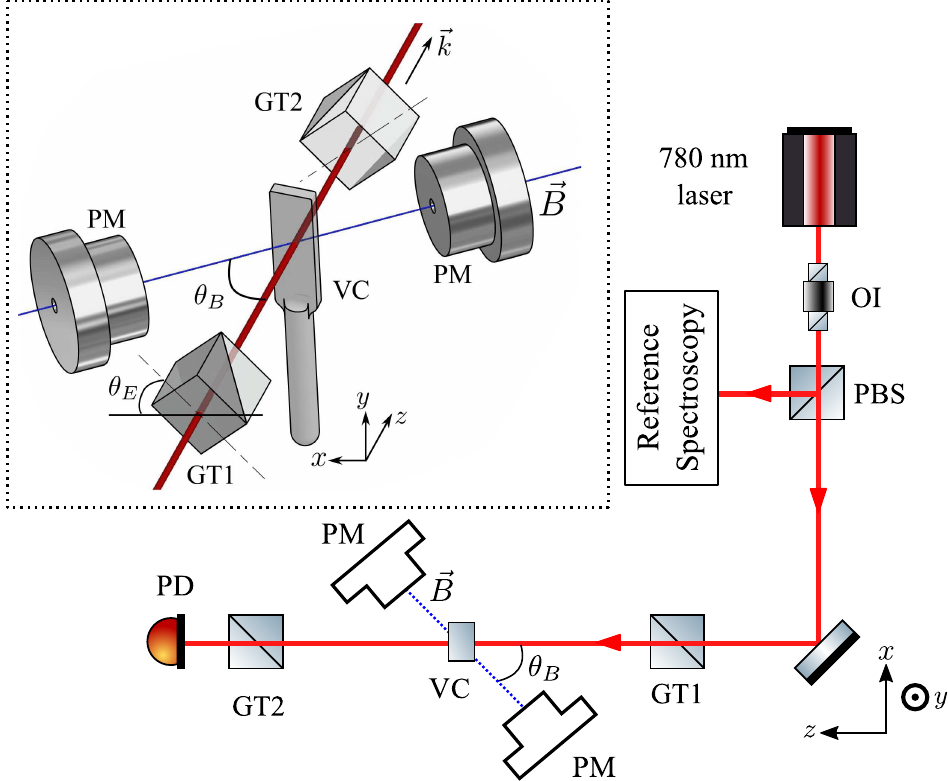}
\caption{Experimental setup for the atomic filter. 780~nm laser light is split on a PBS and one beam is passed through a Rb vapor cell to provide a frequency reference, as in Fig.~\ref{fig:etalonsetup}. The second beam passes through an input Glan-Taylor polarizer (GT1) angled at $\theta_E$ (a variable parameter) to the horizontal $x$-$z$ plane, then through a 5 mm natural abundance Rb vapor cell (VC) and an output polarizer (GT2) fixed at \ang{90} to the first, to be detected on a photodiode (PD).  The vapor cell is mounted in a heater, and is placed in a magnetic field formed between two top-hat shaped permanent magnets (PM). The magnetic field strength, $\vec{B}$, is adjusted by altering the separation of the magnets, and can be up to 0.5~T. The magnets are mounted on a rotation stage so $\theta_B$, the angle between the light propagation direction and the magnetic field direction, can be varied. The inset figure is reproduced with permission from \cite{Keaveney:18}. }
\label{fig:faraday-setup}
\end{figure}

The experimental setup for the atomic filter is shown in Fig.~\ref{fig:faraday-setup}. The beam passes through an input Glan Taylor polarizer angled at $\theta_E$ to the horizontal, then through a 5\,mm natural abundance Rb vapor cell, then an output polarizer which is fixed at \ang{90} to the first. The input polarization angle is a parameter which can be varied. The vapor cell must be short enough to ensure the magnetic field is homogeneous over its length. However decreasing cell length reduces optical depth, though this can be compensated for by increasing the atomic number density by raising temperature \cite{Zentile2015}. Zentile et al. \cite{Zentile_2015_broadening} showed that filter performance degrades as cell length is reduced below a few millimeters because of the extra broadening associated with dipole-dipole interactions at higher densities. The vapor cell is mounted in a heater, and is placed in a magnetic field formed between two top-hat shaped permanent magnets. The magnetic field strength, $\vec{B}$, is adjusted by altering the separation of the magnets, and can produce a maximum field of 0.5 T. The magnets are mounted on a rotation stage so $\theta_B$, the angle between the light propagation direction and the magnetic field direction, can be varied. 

\begin{figure}[htbp]
\centering
\includegraphics[width=\linewidth]{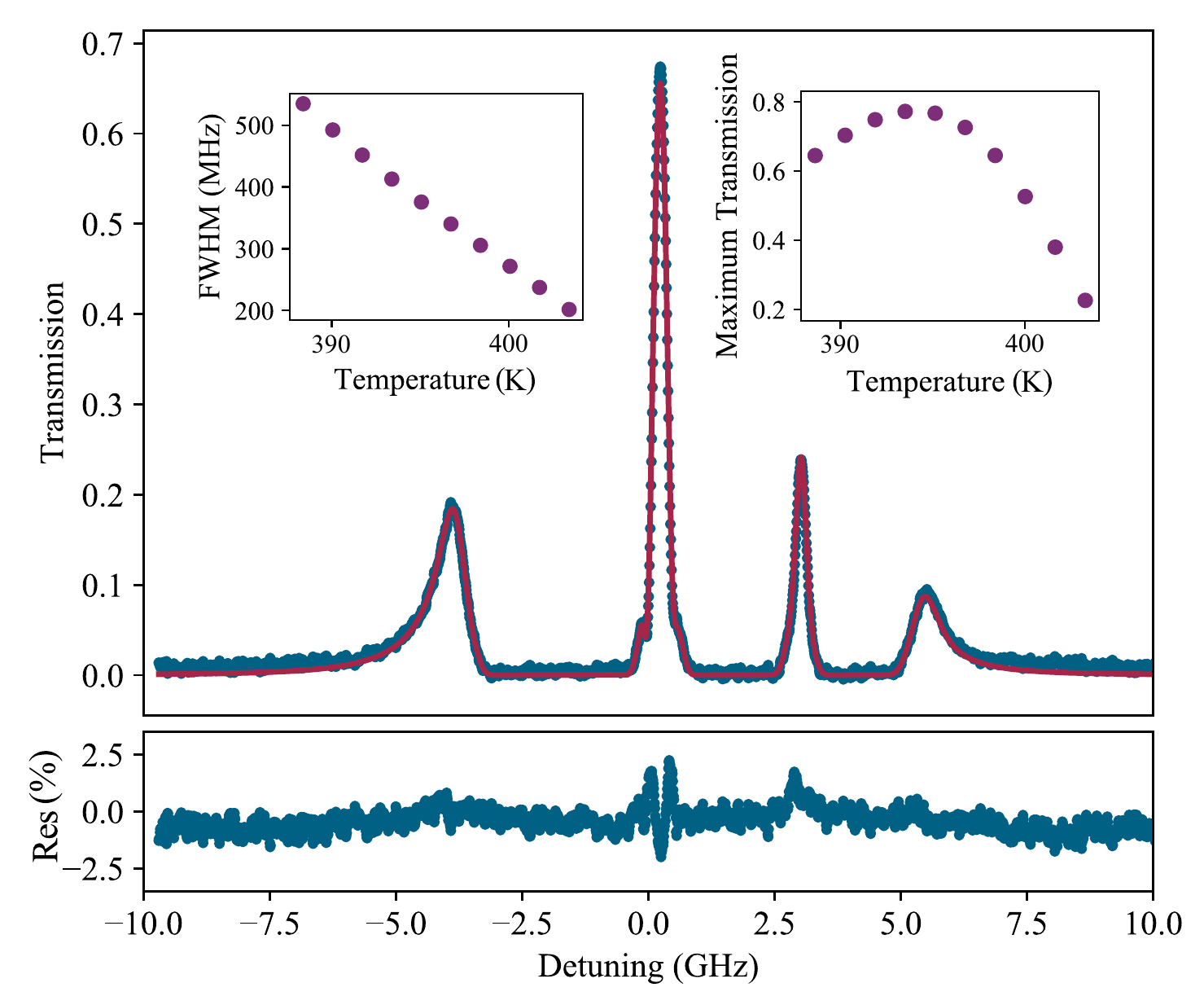}
\caption{Experimental data (blue points) and theoretical fit (red line) for an atomic filter spectrum on Rb D$_2$ line, with fit parameters $T$\,=\,\SI{399.0}{\kelvin}, $|B|$\,=\,218 G, $\theta_B$\,=\,\ang{80.2}, $\theta_E$\,=\,\ang{1.6}. Residuals are displayed and show excellent agreement between theory and experiment. This spectrum has a FWHM of (310\,$\pm$\,1)\,MHz, a maximum transmission of (66\,$\pm$\,1)\,\%, and a FOM of (0.66\,$\pm$\,0.01)\,GHz$^{-1}$. Insets show the effect of changing temperature on the FWHM and maximum transmission of the filter peak.}
\label{fig:faraday-spectrum}
\end{figure}

An atomic filter spectrum with experimental data, ElecSus \cite{Zentile2015} fit and residuals is displayed in Fig.~\ref{fig:faraday-spectrum}. Atomic filters can be broadly classified as two types---line center and wing---depending on where the transmission is relative to the atomic resonance. This filter uses a natural abundance Rb vapor cell (72.17\% $^{85}\text{Rb}$, 27.83\% $^{87}\text{Rb}$) which has narrow peaks at the line centers of the two isotopes, with the peak close to 0\,GHz from the stronger $^{85}\text{Rb}$ transitions. The two outer peaks in the filter spectrum are residual wing-like features \cite{Zentile2015}. To determine the 100\% transmission level, the second polarizer is removed, allowing all of the far off-resonance, unrotated light to pass through the filter. 

This spectrum has a FOM of (0.66\,$\pm$\,0.01)\,GHz$^{-1}$, however optimizing parameters for FOM rather than maximum transmission leads to a filter with a FOM of (1.04\,$\pm$\,0.01)\,GHz$^{-1}$. Changing the temperature, which in turn alters number density, from the value for maximum FOM lowers the FOM, with a steeper drop-off as temperature is increased. The FOM drops to half its maximum value when $T$ is increased by 7\,K or decreased by 10\,K. Transmission characteristics change with all of the fit parameters. Once set, the angles and magnetic field are constant, however temperature will fluctuate. We model the effect of temperature change on the transmission spectrum, and find that while the central peak frequency shift is negligible  ($\sim$10~MHz over \SI{20}{\kelvin}), the peak height and FWHM (Fig.~\ref{fig:faraday-spectrum} insets), and subsidiary peak transmission vary significantly. However temperature fluctuations on the scale expected in the laboratory ($\sim$\SI{1}{\kelvin}) only cause small changes in the transmission spectra (FWHM\,$\sim$\,20\,MHz, $\mathcal{T}_{\text{max}}$\,$\sim$\,0.3\,\%). The FWHM decreases with increasing temperature as the transmission peak is the gap between two absorption features, each of which gets wider as the temperature increases.

\section{Comparison between filter types}

The choice of filter type depends on the requirements of a particular experiment; here we outline how the tested filters compare over a range of criteria, summarized in Table \ref{tab:merits_summary}. 

The cavity filter can have a smaller bandwidth, tens rather than hundreds of MHz, however this is fixed at manufacture while the atomic filter bandwidth can be changed by adjusting experimental parameters. We find a higher maximum transmission for the atomic filter (75\,\%) than for the cavity filter (50\,\%), however the atomic filter has a worse extinction ratio due to having larger subsidiary peaks, and a lower FOM. The extinction ratio and bandwidth of the cavity filter are independent, unlike the atomic filter. When designing the cavity filter, a length can be chosen to produce high extinction at a given frequency from the transmission peak, whereas the frequencies of subsidiary transmission maxima of the atomic filter are fixed. Cavity peaks repeat every FSR, so while extinction of the atomic filter is poorer over 20\,GHz around the main peak, it is much better elsewhere. It has been shown, however, that two cascaded monolithic cavity filters can produce a filter with an effective FSR of hundreds of GHz \cite{Ahlrichs:13}. The FOM calculated across 1\,FSR of this cascaded filter would be lower than that of the original, because the maximum transmission would approximately halve, though the exact numbers would depend on the specific implementation.

The atomic filter is stable to changes in temperature: a change of \SI{1}{\kelvin} negligibly affects transmission frequency ($\sim$\,1\,MHz) and does not alter bandwidth or maximum transmission significantly. In contrast, the cavity filter is very sensitive to temperature: a \SI{1}{\kelvin} change shifts the peak by order 1\,GHz. Conversely this means the cavity filter can be arbitrarily tuned with no change in bandwidth or transmission, while the atomic filter transmits at a fixed frequency determined by the resonances of the atom. 

Imaging is possible through the atomic filter; this is not true for the cavity filter, because it filters spatially, transmitting different cavity modes at different frequencies (Fig.~\ref{fig:fsr-modes}). This property is crucial to applications such as solar imaging.

The atomic filter is very sensitive to the polarization of the input light, which should be linear, and set to match the angle of the first GT polarizer. If this is not the case the transmission will be significantly reduced. Atomic filters can also be used as dichroic beam splitters \cite{Abel:09}. The cavity filter is polarization independent.
\begin{table}[htbp]
\centering
\caption{\bf Summary of relative merits of Lens Cavity and Atomic Line filters}
\begin{tabular}{ccc}
\hline
 & Lens Cavity & Atomic Line \\
\hline
Bandwidth & 10s MHz & 100s MHz \\
On-peak Transmission & 50\,\% & 75\,\% \\
Temperature Stability & 1 mK & 1 K \\
Bandwidth Tunability & Fixed on manufacture & Yes \\
Central Frequency & Arbitrary & Fixed \\
Imaging & No & Yes \\
Polarization & Any & Highly sensitive \\
Footprint & $\sim$\SI{50}{\cm\squared} & $\sim$\SI{2500}{\cm\squared} \\

\hline
\end{tabular}
  \label{tab:merits_summary}
\end{table}

The atomic filter setup used here has a bench footprint of at least 50\,cm\,$\times$\,50\,cm, which is dominated by the rotating stage for the magnets.  This has not been optimized to be as small as possible and could be reduced if the field across the cell is allowed to be non-uniform. The magnet setup is custom made, as is the cell heater, and a 5~mm Rb (or other suitable atomic vapor) cell is required. Our vapor cell was filled in-house, however similar cells are commercially available. Glan-Taylor polarizers are also necessary for maximal extinction of light at undesired frequencies. The cavity filter is experimentally simple, requiring only the temperature stabilized high-reflection coated lens and mode-matching and collimation lenses, and is much smaller with a 7\,cm\,$\times$\,7\,cm footprint. If required, the focal length of the mode-matching lens can be minimized by suitable choice of cavity radius of curvature. The temperature controller is not included in this footprint as it can be mounted away from the setup if long connecting wires are used. 

We conclude that both designs have merits; the atomic filter is effective when signal light is close to an atomic resonance, while the cavity filter allows greater control over bandwidth and is arbitrarily tunable, but requires temperature stabilization.

\noindent\textbf{Funding.} EPSRC (EP/R002061/1); Durham University.

\noindent\textbf{Acknowledgements.} The authors thank Francisco Ponciano-Ojeda for assistance with the atomic filter setup.

\noindent\textbf{Disclosures.} The authors declare no conflicts of interest.

\noindent The data presented in this paper are available from DRO, https://doi.org/10.15128/r2sq87bt63x.


\bibliography{refs_w_title}






\end{document}